\documentclass{emulateapj}

\graphicspath{{./}{figures/}}

\received{January 1, 2018}
\revised{January 7, 2018}
\accepted{\today}
\shorttitle{STARE2}
\shortauthors{Bochenek et al.}
\usepackage{gensymb}
\usepackage{float}
\usepackage[bottom]{footmisc}

\usepackage{natbib}
\begin{document}

\title{STARE2: Detecting Fast Radio Bursts in the Milky Way}

\email{cbochenek@astro.caltech.edu}

%[0000-0003-3875-9568]
\author{Christopher D.\ Bochenek}
\affil{Division of Physics, Math, and Astronomy, California Institute of Technology, Pasadena, CA}

\author{Daniel L.\ McKenna}
\affil{Division of Physics, Math, and Astronomy, California Institute of Technology, Pasadena, CA}

\author{Konstantin V.\ Belov}
\affiliation{Jet Propulsion Laboratory, California Institute of Technology, Pasadena, CA}

\author{Jonathon Kocz}
\affil{Division of Physics, Math, and Astronomy, California Institute of Technology, Pasadena, CA}

\author{S.\ R.\ Kulkarni}
\affil{Division of Physics, Math, and Astronomy, California Institute of Technology, Pasadena, CA}

\author{James Lamb}
\affil{Owens Valley Radio Observatory, California Institute of Technology, Big Pine, CA}

\author{Vikram Ravi}
\affil{Division of Physics, Math, and Astronomy, California Institute of Technology, Pasadena, CA}

\author{David Woody}
\affil{Owens Valley Radio Observatory, California Institute of Technology, Big Pine, CA}



\begin{abstract}

There are several unexplored regions of the short-duration radio transient phase space. One such unexplored region is the luminosity gap between giant pulses (from pulsars) and cosmologically located fast radio bursts (FRBs). The Survey for Transient Astronomical Radio Emission 2 (STARE2) is a search for such transients out to 7\,Mpc. STARE2 has a field of view of 3.6\,steradians  and is sensitive to 1\,millisecond transients above $\sim300$\,kJy. With a two-station system we have detected and localized a solar burst, demonstrating that the pilot system is capable of detecting short duration radio transients. We found no convincing transients with duration between 65\,$\mu$s and 34\,ms in 200\,days of observing, limiting with 95\% confidence the all-sky rate of transients above $\sim300$\,kJy to $< 40\,{\rm sky^{-1}\,year^{-1}}$. If the luminosity function of FRBs could be extrapolated down to 300\,kJy for a distance of 10\,kpc, then one would expect the rate to be $\sim2{\rm\,sky^{-1}\,year^{-1}}$.

\end{abstract}

%%\keywords{editorials, notices --- 
%%miscellaneous --- catalogs --- surveys}

\section{Introduction} \label{sec:intro}
%%Fast radio bursts
%%Low Energy FRBs
%%Previous STARE, Switzerland experiment, how we're different
Just 13 years ago, \citet{Lorimeretal2007} discovered the
first fast radio burst (FRB) that has led to a great amount of activity to find and characterise the population \citep[e.~g.~][]{Championetal2016,Bailesetal2017,Amirietal2018,Koczetal2019, Macquartetal2010}. It is now established that FRBs herald from cosmological distances \citep{Chatterjeeetal2017,Ravi2019,Bannisteretal2019, Prochaskaetal2019} and are of great luminosity.

\begin{figure}
    \centering
    \includegraphics[width = 0.99\linewidth]{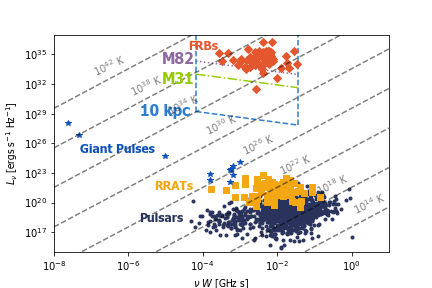}
    \caption{Luminosity of radio transients vs.\ the frequency of the transient times the duration of the transient \citep{Pietkaetal2015}. The blue box shows STARE2's sensitivity to transients within 10\,kpc, the green dot-dashed line shows STARE2's sensitivity to transients from M31, and the purple dotted line shows STARE2's sensitivity to transients from M82. STARE2 is able to detect transients in the gap between pulsars and FRBs, and most FRBs at the distance of M82.}
    \label{fig:phase_space}
\end{figure}

The existence of FRBs shows that processes in nature can produce brightness temperatures of up to $10^{36}$\,K with long enough duration to be seen at cosmological distances. However, as shown in Figure \ref{fig:phase_space}, there is unexplored parameter space between giant pulses from pulsars and FRBs, particularly with respect to luminosity. 

There have been several surveys for extremely bright fast radio transients that may have been sensitive to events in this luminosity gap. STARE \citep{Katzetal2003} was a network of three feeds located throughout the Northeast United States. STARE used a multi-station coincidence approach to filter out radio frequency interference (RFI). Only events seen at all three stations were considered candidates. The experiment ran for 18 months with observations conducted in the 600--613\,MHz band with a time resolution of 0.125\,s and typical flux density threshold at zenith of 27\,kJy. STARE found 3,898\,coincident events and associated all of them with solar radio bursts. 

\citet{Hilaireetal2014} conducted a similar experiment at the Bleien Observatory (near Zurich, Switzerland). It consisted of a 70$\degree$x110$\degree$ FWHM log-periodic antenna  and a 10$\degree$ FWHM horn, both operating in the 1170--1740\,MHz band and a time resolution of 10\,ms and a frequency resolution of 1.02\,MHz. They observed for 289 days with the log-periodic antenna, and 609 days using a horn antenna, achieving a sensitivity to events greater than 200\,kJy for signals 10\,ms in duration. In all, five events were detected. Four happened during the daytime and could have been ``perytons" \citep{Petroffetal2015} or solar radio bursts. The fifth event happened at night with the full moon in the beam of the antenna. \citet{Hilaireetal2014} speculate that this event could have been a solar radio burst reflected off the moon, as its spectrum is similar to that of solar radio bursts. Furthermore, the pulse did not follow the expected $\nu^{-2}$ frequency sweep expected from traveling through the interstellar medium.

Using the fact that they detected no non-solar fast transients, we conclude from their observations with the log-periodic antenna that the all-sky rate of ``intermediate luminosity" fast radio transients is $<15.1\,{\rm sky^{-1}\,yr^{-1}}$ for transients above 400\,kJy and longer than 10\,ms using the statistical framework of \citet{Gehrels1986}. 

Current FRB searches will likely find it difficult to detect nearby fast radio transients in this phase space. If a transient in this luminosity gap was to occur within the Milky Way, it would be a tremendously bright event. So bright, that a dish is not needed to see it. Furthermore, if it did occur in the primary beam of a deep single pulse search, it would likely saturate the instrument and be removed as radio frequency interference (RFI). It is possible that a deep single pulse search would detect such an event in a far sidelobe. For a single dish or beamforming search, it would be difficult to identify that the burst occurred in a far sidelobe without some identifying feature, such as it having the same dispersion measure (DM) as a known pulsar. This would make the burst difficult to place in the luminosity gap and is more likely to be interpreted as a single pulse from a galactic pulsar. However, if baseband data is saved, then it would be possible to determine that the pulse came from far outside of the primary beam. For an image-plane search, one would have to ensure that complex structure in the primary beams of the antennas did not impart a significant amount of noise to the phases and the delay beam did not suppress significantly off-axis emission. Even then, it would be difficult to determine the burst's flux density to even an order of magnitude, as far sidelobes are typically not well characterized. This uncertainty would make it difficult to identify the burst as belonging to this luminosity gap.

Motivated by the considerations discussed above we initiated 
a similar search called the Survey for Transient Astronomical Radio Emission 2 (STARE2). In a nutshell, STARE2 aims to survey the transient radio sky for millisecond  bursts in the 1280--1530\,MHz band. The experiment consists of a single feed pointed at zenith at several sites. Temporal coincidence is used to identify and remove RFI. Our basic time and frequency resolution are 65.536\,$\mu$s and 122\,kHz. We are sensitive to signals of 1\,ms duration above 300\,kJy.

The paper is organized as follows: 
In Section \ref{sec:sources}, we consider different types of sources that STARE2 might be sensitive to. The instrumentation and data processing of STARE2 is described in Section \ref{sec:instrument}. We have commensally detected a solar burst, which is described in Section \ref{sec:burst}. In Section \ref{sec:discussion}, we set an upper limit on the all-sky rate of extremely bright fast radio transients and discuss our future plans.

\section{Extremely Bright Radio Bursts in the Milky Way}
\label{sec:sources}
%%Figure: FRB Energy function
%%Figure: Energy function index limit
%%GRBs and SGRs

Currently, there is no known source of fast transient radio emission beyond the solar system that is bright enough to be seen by STARE2. However, the discovery of fast radio bursts has opened up a vast phase space ranging from Galactic giant radio pulses to cosmologically located FRBs. 

There is a wide range in luminosity of the classical FRBs themselves. It is reasonable to hypothesize that there is a population of FRBs that is too faint to be seen at cosmological distances, but more numerous than classical FRBs such that we should expect the Milky Way to host such a source \citep{Ravi2019}. Even a burst of luminosity $10^{29}$\,ergs\,s$^{-1}$\,Hz$^{-1}$, approximately three orders of magnitude fainter than the lowest luminosity FRB, would be detected at a distance of 10\,kpc with flux density 800\,kJy. Such FRBs, should they exist, would be detected with  could be seen by a simple dipole.

The field of gamma-ray bursts (GRBs) provides another inspiring motivation for the proposed search program \citep[see][]{Kulkarni2018}. There are actually four distinct bursting sources in the gamma-ray sky: terrestrial lightning, soft gamma-ray repeaters (SGRs; these are seen from sources in our Galaxy and the galaxies within a few Mpc of the Sun), short hard bursts (whose typical red-shift is 0.5; associated with neutron star coalescence) and long duration GRBs (whose typical red-shift is 2; associated with deaths of a certain class of massive stars).

In fact, even just over three decades ago SGRs were not seen as distinct from GRBs (either short or hard). We now know that SGRs, as implied by their name repeat and furthermore have a distinctly different origin than that of short or long GRBs. Furthermore, SGRs have moderate luminosity but have a higher volumetric rate than GRBs, as it is rare for a GRB to occur in any individual galaxy in a given year, whereas it is common for an individual galaxy to host multiple SGRs.

This analogy, if applied to FRBs, would suggest that there may be a class of radio bursts which are not as luminous as FRBs but have a much higher volumetric rate than FRBs. Just as with SGRs, these hypothesized sources would be relatively common in the Milky Way.

\subsection{Undiscovered Fast Radio Transients}

In this section, we connect the observed galactic rate of an unknown transient to the volumetric rate, $\Phi(>E)$, of that transient in order to understand what types of events might be seen by STARE2.

If a radio transient tracks stellar mass, then the rate of that transient for a given galaxy is given by Equation \ref{eq:frb_unit_mass}, where $M_{\rm galaxy}$ is the stellar mass of the galaxy and $\Phi_{\rm M}$ is the volumetric rate of stellar mass. We take $\Phi_{\rm M}$ to be $7.4 \times 10^8$\,M$_\odot$\,Mpc$^{-3}$ \citep{Karachentsevetal2018}. Thus, if a radio transient happens once every year in the Milky Way and is detected by STARE2, then its volumetric rate is of order $10^{7}$\,Gpc$^{-3}$\,yr$^{-1}$ above an energy of $2 \times 10^{28}$\,erg\,s$^{-1}$\,Hz$^{-1}$,

\begin{equation}
    R_{\rm galaxy}(>E) = \frac{M_{\rm galaxy}}{\Phi_{\rm M}}  \Phi(>E).
    \label{eq:frb_unit_mass}
\end{equation}

If a radio transient instead tracks star formation instead of stellar mass, then we can similarly relate the volumetric rate of the transient to the star formation rate as shown in Equation \ref{eq:frb_stellar_mass}, where $\Phi_{\rm SFR}$ is the local volumetric rate of star formation and $S$ is the star formation rate for a particular galaxy,

\begin{equation}
    R_{\rm galaxy}(>E) = \frac{S}{\Phi_{\rm SFR}} \Phi(>E).
    \label{eq:frb_stellar_mass}
\end{equation}

\noindent The volumetric star formation rate is $1.95 \times 10^{-2}$\,M$_\odot$\,Mpc$^{-3}$\,yr$^{-1}$ \citep{Salimetal2007}. For a transient with an all-sky rate of 1\,yr$^{-1}$ in the Milky Way, which has a star formation rate of approximately 1\,M$_{\odot}$\,yr$^{-1}$, the volumetric rate is $2 \times 10^{7}$\,Gpc$^{-3}$\,yr$^{-1}$ above $2 \times 10^{28}$\,erg\,s$^{-1}$\,Hz$^{-1}$.

There is approximately 45.8\,M$_\odot$\,yr$^{-1}$ within 3.6\,Mpc, mostly from M82, which has a SFR of 44\,M$_\odot$\,yr$^{-1}$ (Kennicutt \& de los Reyes, 2020 \textit{in prep}). This SFR is calculated using the sum of H$\alpha$ luminosities from all galaxies in \citet{Kennicuttetal2008} with declinations greater than $-30\degree$, excepting M82, whose H$\alpha$ luminosity severely underestimates the SFR. For a transient that has an all-sky rate of $1 {\rm sky^{-1}\,yr^{-1}}$ at the distance of M82, the volumetric rate of fast transients above $5 \times 10^{33}$\,erg\,s$^{-1}$\,Hz$^{-1}$ is approximately $4.3 \times 10^5$\,Gpc$^{-3}$\,yr$^{-1}$.

From this analysis, we see that while STARE2 is sensitive to transients that are bright (have high flux densities) and rare (have low all-sky rates), any transient seen would be relatively faint (have luminosities significantly lower than FRBs) and common (have a high volumetric rate).

\subsection{Low Energy Fast Radio Bursts}

In addition to finding new classes of fast radio transients, it is possible that STARE2 could see low energy FRBs in the Milky Way if the luminosity function of FRBs extends a few orders of magnitude towards lower energies. It is, however, difficult to make a prediction for the rate of Galactic FRBs given the considerable amount of uncertainty in the FRB luminosity function.

\citet{Luetal2019} analyzed a sample of FRBs and found that the number of FRBs above a given energy, $N_{\rm FRB}(>E)$, follows a power law distribution of index $-0.7$, with a cutoff above $\sim 10^{34}$\,ergs\,Hz$^{-1}$. \citet{Luetal2019} also constrained the volumetric rate of FRBs to $>1.1 \times 10^3$\,Gpc$^{-3}$\,yr$^{-1}$ above $10^{32}$\,ergs\,Hz$^{-1}$. We note this rate is a lower limit due to incompleteness of the ASKAP sample and contributions to the DM from the host galaxy and circumgalactic medium of the Milky Way that were not taken into account. \citet{Luoetal2018} analyzed a smaller sample of FRBs and constrained this power law index to between $-0.2$ and $-0.8$. Notably, this is the same energy distribution as exhibited by the first
repeating event, FRB 121102 \citep{Lawetal2017}. However, \citet{Gourdjietal2019} analyzed a sample of low energy FRBs from FRB 121102, and found a much steeper slope of $-1.8$ for energies above $2 \times 10^{37}$\,erg. \citet{Gourdjietal2019} attempt to explain this discrepancy by pointing to the fact that they may have unavoidably underestimated the burst energies, propagation effects \citep{Cordesetal2017}, and by suggesting that a single power-law may not explain the intrinsic energy function from FRB 121102. 

Current observations are not sensitive to the minimum energy of FRBs. \citet{Luoetal2018} set an upper limit on the lower cut-off of the FRB luminosity function of $10^{42}$\,erg\,s$^{-1}$. However, as shown by \citet{Gourdjietal2019}, the FRB luminosity may not be a simple power law with a cut-off. 

%STARE2 can inform us about the low end of the FRB luminosity function. If the FRB luminosity function is steep and can be extended to low energies, then you would expect FRBs to occur in the Milky Way frequently. Alternatively, it may not be valid to extend the FRB luminosity function to low energies as there could be a minimum energy for FRBs.

Using equations \ref{eq:frb_unit_mass} and \ref{eq:frb_stellar_mass}, assuming a volumetric rate of FRBs of $10^{3}$\,Gpc$^{-3}$\,yr$^{-1}$ above $10^{35}$\,erg\,s$^{-1}$\,Hz$^{-1}$, and assuming the luminosity function given by \citet{Luetal2019}, we can infer how far down the luminosity function we must extrapolate such that the rate of FRBs in the Milky Way is $\sim 1$\,yr$^{-1}$. We find that regardless of whether FRBs track stellar mass or star formation, the FRB rate in the Milky Way should be above $\sim 1$\,yr$^{-1}$ for FRBs with energy greater than $\sim 10^{29} \Phi_{3}^{1.4}$\,erg\,s$^{-1}$\,Hz$^{-1}$, where $\Phi_3$ is $\frac{\Phi_{{\rm FRB}}(>10^{35} {\rm\,ergs\,}{\rm s}^{-1}\rm{\,Hz}^-1)}{10^4{\rm\,Gpc\,}^{-3}{\rm \,yr}^{-1}}$. This is approximately 5 orders of magnitude fainter than a typical FRB of luminosity $10^{34}$\,erg\,s$^{-1}$\,Hz$^{-1}$, as defined by Figure \ref{fig:phase_space}. For a distance of 10\,kpc, a luminosity of $4 \times 10^{28}$\,erg\,s$^{-1}$\,Hz$^{-1}$ corresponds to STARE2's sensitivity limit of $\sim300$\,kJy. If the luminosity function of FRBs could be extrapolated down to this luminosity, then one would expect the galactic rate to be $\sim 2{\rm\,year^{-1}}$.

\begin{figure}
    \centering
    \includegraphics[width = 0.99\linewidth]{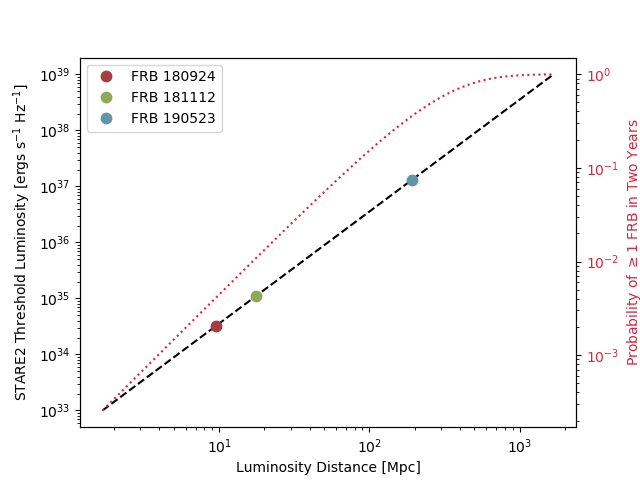}
    \caption{The left axis shows the minimum luminosity of a 1\,ms FRB that would be above STARE2's detection threshold as a function of luminosity distance. The points show the horizon to non-repeating FRBs of known luminosity. The right hand axis shows the probability of at least one FRB occurring above STARE2's detection threshold in a 2 year period as a function of luminosity distance, calculated using the rate and luminosity function given in \citet{Luetal2019}, assuming no maximum energy of FRBs.}
    \label{fig:horizon} \citep{Bannisteretal2019,Prochaskaetal2019,Ravietal2019}
\end{figure}

In addition to low energy FRBs, there is a small chance of finding an FRB with energy equivalent to those FRBs seen at extragalactic distance. STARE2 will be able to detect 1 ms radio transients above $\sim$300\,kJy. Given this, STARE2 could detect an FRB with the same energy as FRB 190523 out to a distance of 191 Mpc \citep{Ravietal2019}. Figure \ref{fig:horizon} shows STARE2's horizon to FRBs for FRBs of different luminosity, as well as the probability of an FRB detectable to STARE2 occurring in a two year period as a function of horizon. Given the  volumetric rate of $\sim 10^{3}$\,Gpc$^{-3}$\,yr$^{-1}$ above an energy of $10^{35}$\,erg\,s$^{-1}$\,Hz$^{-1}$ and luminosity function of \citet{Luetal2019} without a cutoff energy, this implies the probability of an FRB detectable to STARE2 occurring within 191\,Mpc is $\sim 33\%$.

\label{sec:instrument}

 %To probe deeper into the faint end of the luminosity function, we sacrifice sensitivity for field of view in order to find brighter, nearby FRBs. 
%To understand the feasibility of this concept, let us attempt to understand what the FRB population of the Milky Way might look like given the volumetric rate of FRBs with energy greater than $E$, $\Phi_{FRB}(>E)$.

\section{STARE2: The Instrument}
\label{sec:instrument}
\begin{figure}
    \centering
    \includegraphics[width = 0.75\linewidth]{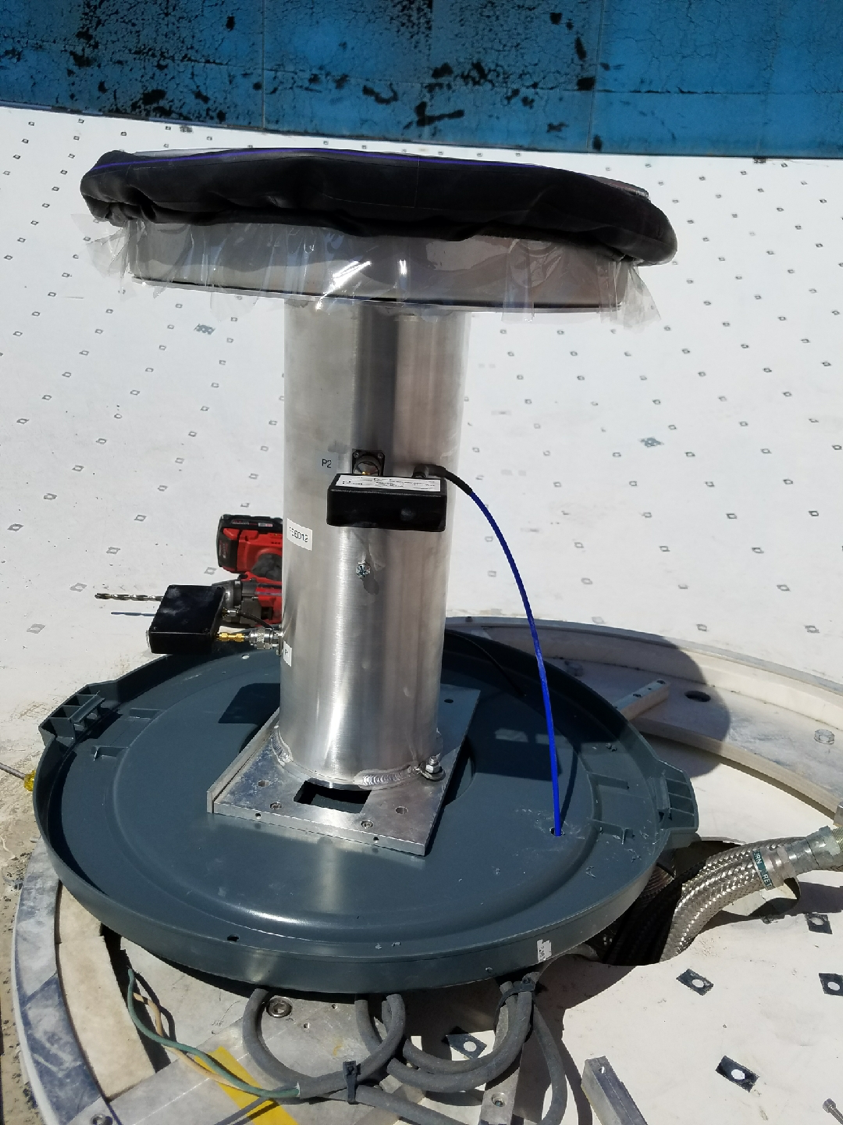}
    \caption{The choke-ring feed located at OVRO. It lies in a 6\,m dish which is used as a horizon shield. The 6\,m dish blocks up to 25$\degree$ above the horizon.}
    \label{fig:feed}
\end{figure}
\begin{figure}
    \centering
    \includegraphics[width = 0.99\linewidth]{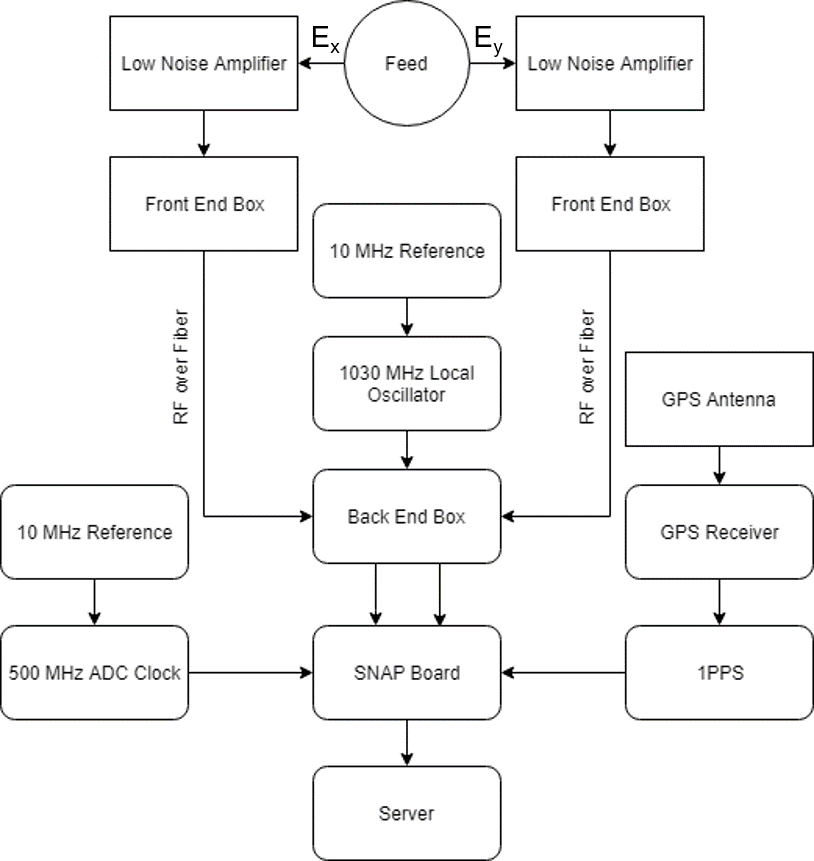}
    \caption{Block diagram for STARE2's system. The blocks with sharp edges are outside in the field and the blocks with rounded edges, except for the feed, are indoors. For a description of the front end box and back end box, see \citep{Koczetal2019}}
    \label{fig:block_diagram}
\end{figure}
\begin{figure}
    \centering
    \includegraphics[width = 0.99\linewidth]{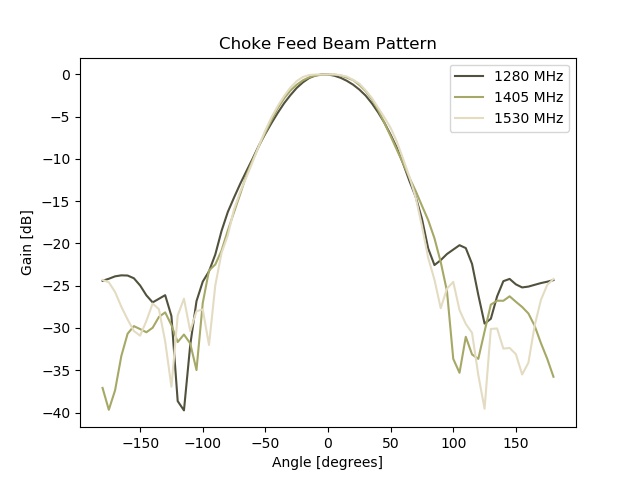}
    \caption{Beam pattern measurement for STARE2's choke ring feed. The FWHM of the beam is $70\degree \pm 5\degree$.}
    \label{fig:beam_pattern}
\end{figure}

STARE2 consists of two dual-polarization choke-ring feeds pointed at zenith at the Owens Valley Radio Observatory (OVRO) and the Venus Antenna at the Goldstone Deep Space Communications Complex (GDSCC). The feed from OVRO is shown in Figure \ref{fig:feed}. A block diagram of the signal path for one station is shown in Figure \ref{fig:block_diagram}. The signal enters the feed through and excites two orthogonally polarized linear probes. We measured the beam pattern of the choke feed by radiating tones ranging from 1165\,MHz to 1665\,MHz. Each tone was generated with the same amount of power, and the distance between the radiating antenna and our feed kept constant at approximately 10\,m. We then measured the power received at each frequency relative to the radiated power with a vector network analyzer. We repeated this measurement for all angles between $-$180$\degree$ and +180$\degree$ at 5$\degree$ increments. There was a concern about reflections off a low wall near the feed, and we placed absorber along this wall to mitigate this concern. The beam pattern of the feed is shown in Figure \ref{fig:beam_pattern}. The FWHM of the beam is $70 \pm 5\degree$.

Each polarization is then amplified by 31 dB across the STARE2 passband by separate low noise amplifiers and travels down coax cables to a front end box. The front end box has a bandpass filter that sets the STARE passband from 1280 MHz--1530 MHz. The signal is then amplified again by 18\,dB and is mixed with an infrared laser driver so that the signal can be sent via optical fiber from the field to a server room. When the signal arrives at the back end box in the server room, the infrared laser signal is converted back to an RF signal. Inside the back end box, the signal goes through a 1200\,MHz--1600\,MHz filter, is amplified again by 46\,dB, mixed with a 1030\,MHz local oscillator so that the 1280\,MHz--1530\,MHz signal is downconverted to a 250\,MHz--500\,MHz intermediate frequency signal. This signal is amplified further and sent to a SNAP board\footnote{https://casper.ssl.berkeley.edu/wiki/SNAP} to be digitized, channelized, and integrated to the desired time resolution of 65.536\,$\mu$s. The SNAP board then sends a spectrum of 2048 16 bit channels at a frequency resolution of 0.12207\,MHz, every 65.536\,$\mu$s to a server.

The system temperature at each site is $60 \pm 5$\,K. This was measured with the y-factor method by pointing the receiver at zenith and measuring the passband over 10 seconds of recorded data with absorber on and off of the feed. This value represents an average across the band. Given our system temperature, we can compute the system equivalent flux density (SEFD) within the FWHM of the beam pattern. We find the SEFD is $19 \pm 2$\,MJy.

The server receives the data from the SNAP, measures the passband, normalizes the data by the shape of the passband, and filters out RFI. The RFI filtering pipeline replaces single pixels in the dynamic spectrum above an SNR of 20, blocks of width 1.95\,MHz and 3.3\,ms above an SNR of 20, and spectral channels that change by more than an SNR of 10 over 1.6\,s or have significantly higher variance than average with the mean value of all pixels. The threshold for the variance is set empirically to remove no channels under typical RFI conditions and let through an single injected 1\,ms long burst with an SNR of 1000. Typically, 25\% of the band is unusable due to RFI. Because a fast radio burst as bright as the ones seen in other galaxies would likely cause the data to meet these criteria, we also enforce that no more than 730 out of 2048 spectral channels will be replaced at any given time so that a broadband burst will not be completely removed from the data.

We use \texttt{heimdall}\citep{Barsdelletal2012} to search the data out of the RFI pipeline for dispersed signals. \texttt{heimdall} dedisperses the data, convolves the data with several matched filters corresponding to different pulse widths, then computes the signal to noise for each time sample, DM, and pulse width, and finally groups high signal to noise candidates of similar times, widths, and DMs together. We search 1546 DMs between 5\,pc\,cm$^{-3}$ and 3000\,pc\,cm$^{-3}$, with a loss in SNR between DM trials of 25\% for a candidate that is one time integration wide \citep{Levin2012}. We search ten logrithmically spaced pulse widths between 65.535\,$\mu$s and 33.5\,ms. We save every candidate above an SNR of 7.3. This threshold was determined empirically to be as low as possible without producing an overwhelming number of candidates. The fact that we require events to be detected independently at multiple sites allows us to use a lower threshold than other experiments. Therefore, for an effective bandwidth of 188\,MHz (the average available bandwidth) and pulse width of 1\,ms, we are sensitive to events above $314 \pm  26$\,kJy.

To estimate the completeness of our detection pipeline, we injected 5484 broadband pulses with a flat spectral index into data from a single station. We detected 4055 of them. The pulses had log uniformly distributed SNRs ranging from 7.3 to $10^{4}$, log uniformly distributed DMs ranging from 5\,pc\,cm$^{-3}$ to 3000\,pc\,cm$^{-3}$, and log uniformly distributed durations between $0.066$\,ms and $45.340$ ms. Because we run each station independently, our two station coincidences are approximately 55\% complete.

The distance between the two stations is 258\,km, corresponding to a light travel time of 0.86\,ms between the stations. Each station generates candidate events independently. In addition to our single station RFI removal, RFI is filtered out by enforcing that a candidate was found at each site within one light travel time, 0.86\,ms plus the maximum duration of a candidate, 33.554\,ms, which gives a window in time for coincident events of 34.414\,ms. 

In order to successfully detect coincident events at multiple independent sites, keeping accurate absolute time is critical. To time-stamp the start of an observation, we use a GPS receiver that produces one pulse per second (1PPS) at the start of every GPS second. This receiver is locked to a rubidium clock to minimize drift on small timescales. This 1PPS is accurate to within 250\,ns. To keep track of time during an observation, we rely on the clock that runs the ADC. This clock is locked to a 10\,MHz reference that is good to better than 2 parts in 10$^{9}$. We obtain a new 1PPS pulse every 5 hours to reset the system. The difference in cable lengths between the two sites should also impart an offset of no more than 3\,$\mu$s, with the station at OVRO having longer cables.

The RFI filtering and candidate search pipeline produces a candidate rate of 83\,hr$^{-1}$ at OVRO and 270\,hr$^{-1}$ at GDSCC. The rate of coincidences is 5\,day$^{-1}$. This rate is consistent with the expected rate of coincidences assuming that the candidates at OVRO are unrelated to the candidates at GDSCC, and while there are correlated events associated with the Sun and RFI, they are a minority of candidates. Real correlated events are distinguished by their dynamic spectra and time delay between the two sites.

\begin{figure}
    \centering
    \includegraphics[width = 0.99\linewidth]{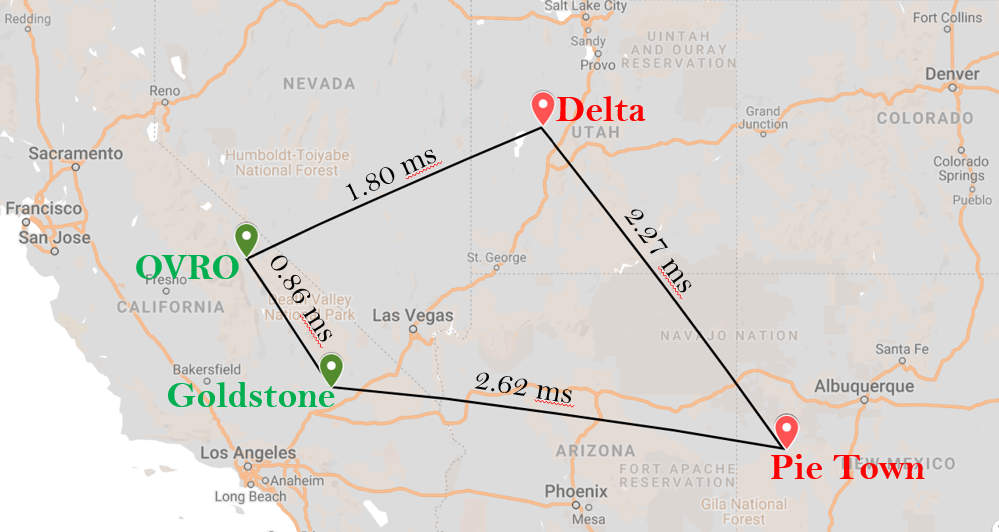}
    \caption{Map of STARE2 stations with light travel time delays between the baselines. The OVRO and Goldstone sites in green are operational, while the Delta and Pie Town sites in red are options for additional stations.}
    \label{fig:map}
\end{figure}

%%Figure: Detector Map
%%Figure: Picture of detectors
%%Figure: STARE system block diagram

%%Figure: Predicted vs. actual time delays from NAVSTAR 77
%%Important numbers: Bandwidth, bandwidth removed by RFI, time resolution, frequency resolution, Tsys of each station, sensitivity to a 1 ms burst, sky coverage, establish timing accuracy to better than 1 microsec, light travel time between each station, DM trials, filter widths, system saturation flux (ADC)
%%Things to describe: Data pipeline from feed, ADC, FPGA, RFI, searching, coincidencing. Beam pattern measurements. Testing with GPS L3 or coincident candidates.

\section{Detection of a Solar Burst}
\label{sec:burst}
\begin{figure*}
    \centering
    \includegraphics[width = 0.99\linewidth]{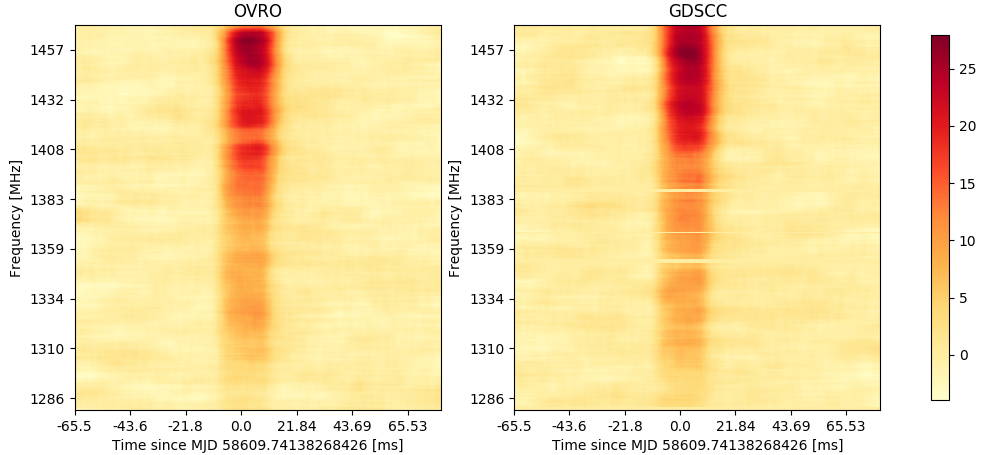}
    \caption{Left: The dynamic spectrum of ST 190506B from OVRO. Right: The dynamic spectrum of ST 190506B from GDSCC. The same solar burst is seen from both OVRO and GDSCC. The data have been corrected for the shape of the passband and normalized to unit variance. They have also been averaged to the width of the burst in time and every 7.8\,MHz in frequency. Channels with significant RFI have been replaced with zero. The colorbar shows the signal to noise in each pixel after averaging.}
    \label{fig:imshow}
\end{figure*}

%SNR 283, width 295 pm 1 samples=19.33312 pm .06 ms, time delay uncertainty 68.3 microsec, time delay: 394 microsec, expected: 404 microseconds, flux: 9.1 pm .8 MJy, beam gain (dB) -.72, sun elevation: 50 degrees MJD: 58609.4497160176 PDT, MJD 58609.7413826842 UTC, UTC: 2019, May 6th 17:47:35.385 UTC

On May 6th, 2019, at UTC 17:47:35.385, we detected a candidate event, ST 190506B, at both OVRO and GDSCC. The naming scheme of the candidates begins with an abbreviation of the instrument, followed by the date, shown as YYMMDD, and a letter or series of letters corresponding to the order of coincidences detected on that day. The dynamic spectrum of this event at each station is shown in Figure \ref{fig:imshow}.

\begin{figure}
    \centering
    \includegraphics[width = 0.99\linewidth]{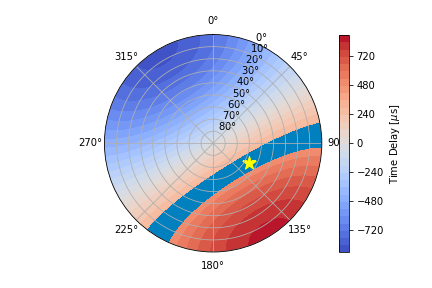}
    \caption{Localization region for ST 190506B in altitude and azimuth at OVRO. The colorbar maps a time delay between each station to a part of the sky. The localization region for ST 190506B is shown in blue. The Sun's position at the time of the burst is shown by the yellow star. The time delay expected from the Sun is within the uncertainty of the measured time delay.}
    \label{fig:localization}
\end{figure}

The measured delay time between OVRO and GDSCC was $394 \pm 68$\,$\mu$s. The expected delay time for the Sun between the two stations at the time of the burst was 404\,$\mu$s. This delay time corresponds to the localization region shown in \ref{fig:localization}, and is consistent with the position of the Sun.

The width of the burst is $19.33 \pm 0.06$\,ms and it has a beam-corrected flux density of $9.1 \pm 0.8$\,MJy. The burst was detected with a DM of 5\,pc\,cm$^{-3}$, however it is consistent with a DM of 0\,pc\,cm$^{-3}$. The burst was also seen during a time of heightened solar activity, and X-ray data from the \textit{GOES} satellite show a significantly higher flux than average at the time of the burst. The burst width, flux density, and solar activity during the time of the burst is consistent with a solar burst \citep{Melendezetal1999, Benzetal1983}. From this event, we conclude that the system can commensally detect FRB-like events of similar flux density, should they occur.
%%Figure: dynamic spectra from OVRO and GDSCC
%%Figure: Altitude and azimuth localization
%%Figure: Delay vs frequency

\section{Discussion}
\label{sec:discussion}
%%Implications for FRB energy function & cutoff luminosity
%%Future stations/improvements
We have been observing for 200 days and have seen no convincing fast transient events. We can therefore set a 95\% confidence upper limit on the all-sky rate of extremely bright fast radio transients of 40\,sky$^{-1}$\,yr$^{-1}$ above $314 \pm 26$\,kJy. %If fast transients track stellar mass, this limits the volumetric rate to $< 1.2 \times 10^{9}$\,Gpc$^{-3}$\,yr$^{-1}$ above $2 \times 10^{28}$\,erg\,s$^{-1}$\,Hz$^{-1}$. If fast radio transients track star formation, this limits the volumetric rate to $< 2.4 \times 10^{9}$\,Gpc$^{-3}$\,yr$^{-1}$ above $2 \times 10^{28}$\,erg\,s$^{-1}$\,Hz$^{-1}$.

\begin{figure}
    \centering
    \includegraphics[width = 0.99\linewidth]{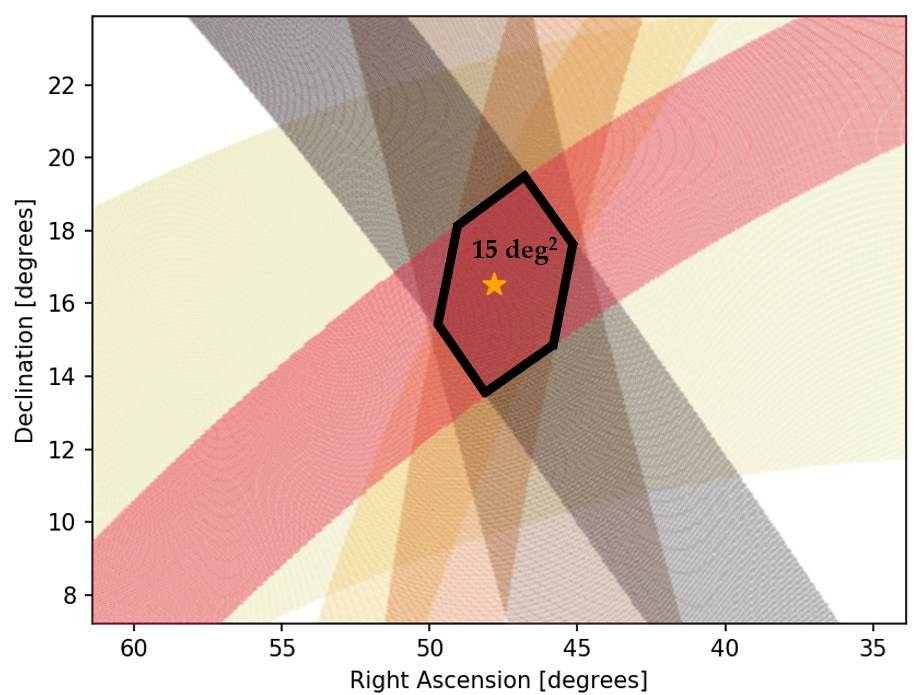}
    \caption{Localization region for ST 190506B in right ascension and declination, assuming the four station system in Figure \ref{fig:map} and an uncertainty of 68\,$\mu$s in delay time. The overlap of the localization arcs is 15\,deg$^2$. The overlap of all the localization arcs is marked by a black hexagon.}
    \label{fig:four_loc}
\end{figure}

We plan to expand our network to four stations as shown in \ref{fig:four_loc}. With four stations, we will be able to localize an event to a patch of sky, rather than a stripe. We will also be able to determine the altitude of a source below approximately $10^3$\,km, in a manner similar to the Global Positioning System. As shown in Figure \ref{fig:four_loc}, with the proposed four station system, we would have localized ST 190506B to 15\,deg$^2$. As of November 2019, the station in Delta, Utah is operational.

\section{Acknowledgements}

We would like to thank the then director of OVRO, A.\ Readhead, for funds (derived from the Alan Moffet Funds) which allowed us to start this project. The Presidential Director's Fund (PDF) allowed us to build the second system at Goldstone. We are thankful to Caltech and the Jet Propulsion Laboratories for the second round of funding. We would also like to thank Sandy Weinreb and David Hodge for building the front end and back end boxes, Jeffrey Lagrange for his support at GDSCC, and the entire OVRO staff for their support. Part of the research was carried out at the Jet Propulsion Laboratory, California Institute of Technology, under a contract with the National Aeronautics and Space Administration.

\bibliographystyle{aasjournal}
\bibliography{bibliography.bib}

\begin{thebibliography}{}
\expandafter\ifx\csname natexlab\endcsname\relax\def\natexlab#1{#1}\fi
\providecommand{\url}[1]{\href{#1}{#1}}

\bibitem[{{Amiri} {et~al.}(2018){Amiri}, {Bandura}, {Berger}, {Bhardwaj},
  {Boyce}, {Boyle}, {Brar}, {Burhanpurkar}, {Chawla}, {Chowdhury}, {Cliche},
  {Cranmer}, {Cubranic}, {Deng}, {Denman}, {Dobbs}, {Fandino}, {Fonseca},
  {Gaensler}, {Giri}, {Gilbert}, {Good}, {Guliani}, {Halpern}, {Hinshaw},
  {H{\"o}fer}, {Josephy}, {Kaspi}, {Landecker}, {Lang}, {Liao}, {Masui},
  {Mena-Parra}, {Naidu}, {Newburgh}, {Ng}, {Patel}, {Pen},
  {Pinsonneault-Marotte}, {Pleunis}, {Rafiei Ravandi}, {Ransom}, {Renard},
  {Scholz}, {Sigurdson}, {Siegel}, {Smith}, {Stairs}, {Tendulkar}, {Vand
  erlinde}, \& {Wiebe}}]{Amirietal2018}
{Amiri}, M., {Bandura}, K., {Berger}, P., {et~al.} 2018, \apj, 863, 48

\bibitem[{{Bailes} {et~al.}(2017){Bailes}, {Jameson}, {Flynn}, {Bateman},
  {Barr}, {Bhandari}, {Bunton}, {Caleb}, {Campbell-Wilson}, {Farah},
  {Gaensler}, {Green}, {Hunstead}, {Jankowski}, {Keane}, {Krishnan}, {Murphy},
  {O'Neill}, {Os{\l}owski}, {Parthasarathy}, {Ravi}, {Rosado}, \&
  {Temby}}]{Bailesetal2017}
{Bailes}, M., {Jameson}, A., {Flynn}, C., {et~al.} 2017, PASA, 34, e045

\bibitem[{{Bannister} {et~al.}(2019){Bannister}, {Deller}, {Phillips},
  {Macquart}, {Prochaska}, {Tejos}, {Ryder}, {Sadler}, {Shannon}, {Simha},
  {Day}, {McQuinn}, {North-Hickey}, {Bhandari}, {Arcus}, {Bennert}, {Burchett},
  {Bouwhuis}, {Dodson}, {Ekers}, {Farah}, {Flynn}, {James}, {Kerr}, {Lenc},
  {Mahony}, {O{\textquoteright}Meara}, {Os{\l}owski}, {Qiu}, {Treu}, {U},
  {Bateman}, {Bock}, {Bolton}, {Brown}, {Bunton}, {Chippendale}, {Cooray},
  {Cornwell}, {Gupta}, {Hayman}, {Kesteven}, {Koribalski}, {MacLeod},
  {McClure-Griffiths}, {Neuhold}, {Norris}, {Pilawa}, {Qiao}, {Reynolds},
  {Roxby}, {Shimwell}, {Voronkov}, \& {Wilson}}]{Bannisteretal2019}
{Bannister}, K.~W., {Deller}, A.~T., {Phillips}, C., {et~al.} 2019, Science,
  365, 565

\bibitem[{{Barsdell} {et~al.}(2012){Barsdell}, {Bailes}, {Barnes}, \&
  {Fluke}}]{Barsdelletal2012}
{Barsdell}, B.~R., {Bailes}, M., {Barnes}, D.~G., \& {Fluke}, C.~J. 2012, in
  Astronomical Society of the Pacific Conference Series, Vol. 461, Astronomical
  Data Analysis Software and Systems XXI, ed. P.~{Ballester}, D.~{Egret}, \&
  N.~P.~F. {Lorente}, 37

\bibitem[{{Benz} {et~al.}(1983){Benz}, {Bernold}, \& {Dennis}}]{Benzetal1983}
{Benz}, A.~O., {Bernold}, T.~E.~X., \& {Dennis}, B.~R. 1983, \apj, 271, 355

\bibitem[{{Champion} {et~al.}(2016){Champion}, {Petroff}, {Kramer}, {Keith},
  {Bailes}, {Barr}, {Bates}, {Bhat}, {Burgay}, {Burke-Spolaor}, {Flynn},
  {Jameson}, {Johnston}, {Ng}, {Levin}, {Possenti}, {Stappers}, {van Straten},
  {Thornton}, {Tiburzi}, \& {Lyne}}]{Championetal2016}
{Champion}, D.~J., {Petroff}, E., {Kramer}, M., {et~al.} 2016, \mnras, 460, L30

\bibitem[{{Chatterjee} {et~al.}(2017){Chatterjee}, {Law}, {Wharton},
  {Burke-Spolaor}, {Hessels}, {Bower}, {Cordes}, {Tendulkar}, {Bassa},
  {Demorest}, {Butler}, {Seymour}, {Scholz}, {Abruzzo}, {Bogdanov}, {Kaspi},
  {Keimpema}, {Lazio}, {Marcote}, {McLaughlin}, {Paragi}, {Ransom}, {Rupen},
  {Spitler}, \& {van Langevelde}}]{Chatterjeeetal2017}
{Chatterjee}, S., {Law}, C.~J., {Wharton}, R.~S., {et~al.} 2017, \nat, 541, 58

\bibitem[{{Cordes} {et~al.}(2017){Cordes}, {Wasserman}, {Hessels}, {Lazio},
  {Chatterjee}, \& {Wharton}}]{Cordesetal2017}
{Cordes}, J.~M., {Wasserman}, I., {Hessels}, J.~W.~T., {et~al.} 2017, \apj,
  842, 35

\bibitem[{{Gehrels}(1986)}]{Gehrels1986}
{Gehrels}, N. 1986, \apj, 303, 336

\bibitem[{{Gourdji} {et~al.}(2019){Gourdji}, {Michilli}, {Spitler}, {Hessels},
  {Seymour}, {Cordes}, \& {Chatterjee}}]{Gourdjietal2019}
{Gourdji}, K., {Michilli}, D., {Spitler}, L.~G., {et~al.} 2019, arXiv e-prints,
  arXiv:1903.02249

\bibitem[{{Karachentsev} \& {Telikova}(2018)}]{Karachentsevetal2018}
{Karachentsev}, I.~D., \& {Telikova}, K.~N. 2018, Astronomische Nachrichten,
  339, 615

\bibitem[{{Katz} {et~al.}(2003){Katz}, {Hewitt}, {Corey}, \&
  {Moore}}]{Katzetal2003}
{Katz}, C.~A., {Hewitt}, J.~N., {Corey}, B.~E., \& {Moore}, C.~B. 2003,
  Publications of the Astronomical Society of the Pacific, 115, 675

\bibitem[{{Kennicutt} {et~al.}(2008){Kennicutt}, {Lee}, {Funes}, {J.}, {Sakai},
  \& {Akiyama}}]{Kennicuttetal2008}
{Kennicutt}, Jr., R.~C., {Lee}, J.~C., {Funes}, J.~G., {et~al.} 2008, \apjs,
  178, 247

\bibitem[{{Kocz} {et~al.}(2019){Kocz}, {Ravi}, {Catha}, {D'Addario},
  {Hallinan}, {Hobbs}, {Kulkarni}, {Shi}, {Vedantham}, {Weinreb}, \&
  {Woody}}]{Koczetal2019}
{Kocz}, J., {Ravi}, V., {Catha}, M., {et~al.} 2019, \mnras, 2146

\bibitem[{{Kulkarni}(2018)}]{Kulkarni2018}
{Kulkarni}, S.~R. 2018, Nature Astronomy, 2, 832

\bibitem[{{Law} {et~al.}(2017){Law}, {Abruzzo}, {Bassa}, {Bower},
  {Burke-Spolaor}, {Butler}, {Cantwell}, {Carey}, {Chatterjee}, {Cordes},
  {Demorest}, {Dowell}, {Fender}, {Gourdji}, {Grainge}, {Hessels}, {Hickish},
  {Kaspi}, {Lazio}, {McLaughlin}, {Michilli}, {Mooley}, {Perrott}, {Ransom},
  {Razavi-Ghods}, {Rupen}, {Scaife}, {Scott}, {Scholz}, {Seymour}, {Spitler},
  {Stovall}, {Tendulkar}, {Titterington}, {Wharton}, \&
  {Williams}}]{Lawetal2017}
{Law}, C.~J., {Abruzzo}, M.~W., {Bassa}, C.~G., {et~al.} 2017, \apj, 850, 76

\bibitem[{{Levin}(2012)}]{Levin2012}
{Levin}, L. 2012, PhD thesis, Swinburne University of Technology

\bibitem[{{Lorimer} {et~al.}(2007){Lorimer}, {Bailes}, {McLaughlin},
  {Narkevic}, \& {Crawford}}]{Lorimeretal2007}
{Lorimer}, D.~R., {Bailes}, M., {McLaughlin}, M.~A., {Narkevic}, D.~J., \&
  {Crawford}, F. 2007, Science, 318, 777

\bibitem[{{Lu} \& {Piro}(2019)}]{Luetal2019}
{Lu}, W., \& {Piro}, A.~L. 2019, \apj, 883, 40

\bibitem[{{Luo} {et~al.}(2018){Luo}, {Lee}, {Lorimer}, \&
  {Zhang}}]{Luoetal2018}
{Luo}, R., {Lee}, K., {Lorimer}, D.~R., \& {Zhang}, B. 2018, \mnras, 481, 2320

\bibitem[{{Macquart} {et~al.}(2010){Macquart}, {Bailes}, {Bhat}, {Bower},
  {Bunton}, {Chatterjee}, {Colegate}, {Cordes}, {D'Addario}, {Deller},
  {Dodson}, {Fender}, {Haines}, {Halll}, {Harris}, {Hotan}, {Johnston},
  {Jones}, {Keith}, {Koay}, {Lazio}, {Majid}, {Murphy}, {Navarro}, {Phillips},
  {Quinn}, {Preston}, {Stansby}, {Stairs}, {Stappers}, {Staveley-Smith},
  {Tingay}, {Thompson}, {van Straten}, {Wagstaff}, {Warren}, {Wayth}, {Wen}, \&
  {CRAFT Collaboration}}]{Macquartetal2010}
{Macquart}, J.-P., {Bailes}, M., {Bhat}, N.~D.~R., {et~al.} 2010, PASA, 27, 272

\bibitem[{{Mel{\'e}ndez} {et~al.}(1999){Mel{\'e}ndez}, {Sawant}, {Fernandes},
  \& {Benz}}]{Melendezetal1999}
{Mel{\'e}ndez}, J.~L., {Sawant}, H.~S., {Fernandes}, F.~C.~R., \& {Benz}, A.~O.
  1999, \solphys, 187, 77

\bibitem[{{Petroff} {et~al.}(2015){Petroff}, {Keane}, {Barr}, {Reynolds},
  {Sarkissian}, {Edwards}, {Stevens}, {Brem}, {Jameson}, {Burke-Spolaor},
  {Johnston}, {Bhat}, {Kudale}, \& {Bhand ari}}]{Petroffetal2015}
{Petroff}, E., {Keane}, E.~F., {Barr}, E.~D., {et~al.} 2015, \mnras, 451, 3933

\bibitem[{{Pietka} {et~al.}(2015){Pietka}, {Fender}, \&
  {Keane}}]{Pietkaetal2015}
{Pietka}, M., {Fender}, R.~P., \& {Keane}, E.~F. 2015, \mnras, 446, 3687

\bibitem[{{Prochaska} {et~al.}(2019){Prochaska}, {Macquart}, {McQuinn},
  {Simha}, {Shannon}, {Day}, {Marnoch}, {Ryder}, {Deller}, {Bannister},
  {Bhandari}, {Bordoloi}, {Bunton}, {Cho}, {Flynn}, {Mahony}, {Phillips},
  {Qiu}, \& {Tejos}}]{Prochaskaetal2019}
{Prochaska}, J.~X., {Macquart}, J.-P., {McQuinn}, M., {et~al.} 2019, arXiv
  e-prints, arXiv:1909.11681

\bibitem[{{Ravi}(2019)}]{Ravi2019}
{Ravi}, V. 2019, arXiv e-prints, arXiv:1907.06619

\bibitem[{{Ravi} {et~al.}(2019){Ravi}, {Catha}, {D'Addario}, {Djorgovski},
  {Hallinan}, {Hobbs}, {Kocz}, {Kulkarni}, {Shi}, {Vedantham}, {Weinreb}, \&
  {Woody}}]{Ravietal2019}
{Ravi}, V., {Catha}, M., {D'Addario}, L., {et~al.} 2019, \nat, 572, 352

\bibitem[{{Saint-Hilaire} {et~al.}(2014){Saint-Hilaire}, {Benz}, \&
  {Monstein}}]{Hilaireetal2014}
{Saint-Hilaire}, P., {Benz}, A.~O., \& {Monstein}, C. 2014, \apj, 795, 19

\bibitem[{{Salim} {et~al.}(2007){Salim}, {Rich}, {Charlot}, {Brinchmann},
  {Johnson}, {Schiminovich}, {Seibert}, {Mallery}, {Heckman}, {Forster},
  {Friedman}, {Martin}, {Morrissey}, {Neff}, {Small}, {Wyder}, {Bianchi},
  {Donas}, {Lee}, {Madore}, {Milliard}, {Szalay}, {Welsh}, \&
  {Yi}}]{Salimetal2007}
{Salim}, S., {Rich}, R.~M., {Charlot}, S., {et~al.} 2007, The Astrophysical
  Journal Supplement Series, 173, 267

\end{thebibliography}

\end{document}